\documentstyle[12pt]{article}
\input epsf

\newcommand{\resection}[1]{\setcounter{equation}{0}\section{#1}}

\def\a{\alpha}

\def\g {\gamma}

\def\t{\theta}

\def\e{\epsilon}

\def\s {\sigma}

\def\M{{\cal M}}

\def\bd{\begin{displaystyle}}
\def\ed{\end{displaystyle}}
\def\ba{\begin{array}}
\def\ea{\end{array}}

\def\be{\begin{equation}}
\def\ee{\end{equation}}
\def\bea{\begin{eqnarray}}
\def\eea{\end{eqnarray}}
\def\beano{\begin{eqnarray*}}
\def\eeano{\end{eqnarray*}}

\def\Jrnl#1#2#3#4{{#1} {\bf #2}, (#4) #3}
\def\NPB{{ Nucl. Phys.} \bf B}
\def\PLB{{ Phys. Lett.}  \bf B}
\def\PRL{ Phys. Rev. Lett.}
\def\PRA{{ Phys. Rev.} \bf A}

\def\IJMPA{{ Int. Jour. Mod. Phys.} \bf A}

\def\AnPh{Ann. Phys. }
\def\SJNP{Sov. J. Nucl. Phys.}
\def\TMP{Theor. Math. Phys. }

\begin{document}
\oddsidemargin 5mm
\setcounter{page}{0}
\newpage     
\setcounter{page}{0}
\begin{titlepage}
\begin{flushright}
IC/98/28\\
SISSA/EP/98/28
\end{flushright}
\vspace{0.5cm}

\begin{center}
\renewcommand{\thefootnote}{\fnsymbol{footnote}}
{\large {\bf Tricritical Ising Model with a Boundary}}
\footnote{Work done under 
partial support of the EC TMR Programme {\em Integrability, 
non--perturbative effects and symmetry in Quantum Field Theories}, grant
FMRX-CT96-0012} \\
\vspace{1.5cm}
{\bf 
A. De Martino} \footnote{\tt{email: demarti@sissa.it}}\\
{\em International School for Advanced Studies, Via Beirut 2-4, 
34014 Trieste, Italy} \\
{\em Istituto Nazionale di Fisica Nucleare, Sezione di Trieste}\\
\vspace{0.8cm}
{\bf 
M. Moriconi} \footnote{\tt{email: moriconi@ictp.trieste.it}}\\
{\em The Abdus Salam International Centre for Theoretical Physics}\\
{\em Strada Costiera 11, 34100 Trieste, Italy}\\
\end{center}
\renewcommand{\thefootnote}{\arabic{footnote}}
\vspace{6mm}

\begin{abstract}
\noindent
We study the integrable and supersymmetric massive $\hat\phi_{(1,3)}$
deformation of the 
tricritical Ising model in the presence of a boundary. 
We use constraints from supersymmetry in order to compute the exact 
boundary $S$-matrices, which turn out to depend explicitly on the topological 
charge of the supersymmetry algebra. 
We also solve the general boundary Yang-Baxter equation and show 
that in appropriate limits the general reflection matrices go over the 
supersymmetry preserving solutions.
Finally, we briefly discuss the possible connection between our
reflection matrices and boundary perturbations
within the framework of  perturbed boundary conformal field theory.

\vspace{3cm}

\end{abstract}
\vspace{5mm}
\end{titlepage}

\newpage
\setcounter{footnote}{0}
\renewcommand{\thefootnote}{\arabic{footnote}}

\resection{Introduction}

There are several problems in different areas of theoretical physics
that involve the study of boundary field theories, such as
the Kondo effect, quantum impurities in strongly correlated electron systems,
the catalysis of baryon decay in the
presence of magnetic monopoles (Callan-Rubakov effect),
and even black-hole evaporation, to name a few.
Therefore the study of boundary theories is more than an interesting exercise,
and we should try to learn as much  as possible about them. 
An especially interesting class of boundary field theories can be obtained 
by restricting 1+1-dimensional integrable field theories to the half-line
while preserving integrability. A remarkable example is the
boundary sine-Gordon model \cite{gzam}, 
which has found very important applications \cite{flsprl} in real physical 
systems in the past few years.

In this paper we study one of the simplest two-dimensional models which
has nonetheless a very rich structure, the tricritical Ising model (TIM),
in the presence of a boundary. 
The TIM provides a useful venue to study many non-trivial aspects of 
two-dimensional
quantum field theory such as superconformal invariance \cite{fqspl,qiu},
renormalization group flows \cite{kms,zamrf} and exact S-matrices 
\cite{zamtim,fendley}. 

As a lattice model the TIM can be realized as an 
Ising model with annealed vacancies \cite{betal}, with Hamiltonian
\be
H= -J \sum_{\langle i,j \rangle} \sigma_i \sigma_j - \mu \sum_i (\sigma_i)^2,
\label{ham}
\ee
where the first sum is performed over nearest neighbors,  
$\sigma_i = \pm 1\ , 0$ are the spin variables, $\mu$ is the chemical 
potential and $J$ is the energy of a configuration of a pair of unlike spins. 
This model has a critical point for some value of $(J,\mu)$ where three 
phases can coexist and to which can be 
associated \cite{fqsprl} a conformal field theory with 
central charge $c=7/10$, corresponding both to the next
simplest minimal model ${\cal M}_4$ and to 
the simplest $N=1$ superconformal minimal model ${\cal{SM}}_3$.
This fact will be extremely useful in the following.

In \cite{zam86} Zamolodchikov has shown that unitary minimal models
can be associated to the infrared fixed point of some particular 
scalar field theories, 
having an effective Landau-Ginzburg (LG) description. 
To the $\M _m$ model, with central charge $c=1-6/m(m+1)$, $m=3,4,\ldots$, 
one can associate the action
\be
S_{LG}=\int d^2 z \, \left[{1 \over 2} (\partial \phi)^2 + 
\phi^{2m-2}\right] \ . \label{lg}
\ee
There is also a LG description for the $N=1$ superconformal 
unitary minimal series ${\cal SM}_n$, with central charge $c=3/2-12/n(n+2)$, 
$n=3,4,\ldots$, given by the action
\be
S_{LG}^{N=1} = \int d^2 z \, d^2 \theta \, \left[{1 \over 2} 
(D \Phi)^2+ \Phi^{n}\right]\ ,
\label{slg}
\ee
where the superfield $\Phi$ written in components is 
$\Phi=\phi+{\mathbf\t} \psi + {\bar{\mathbf\t}} {\bar{\psi}} 
+ {\mathbf\t} {\bar{\mathbf\t}} F$, and the conformal dimensions
for the fields $\phi$, $\psi$, $\bar{\psi}$ and $F$ are $(1/10,1/10)$,
$(3/5,1/10)$, $(1/10,3/5)$ and $(3/5,3/5)$ respectively. 
Therefore the conformal theory associated to the 
TIM can be studied as the critical point of a bosonic theory with a
$\phi^6$ potential or as a $N=1$ supersymmetric theory with a $\Phi^3$
potential. 

Any $N=1$ superconformal field theory (SCFT) allows two different projections
onto local field theories \cite{fqspl}. This
comes about in the following way. The fields in a SCFT 
can be divided in two types, Neveu-Schwarz (NS) and Ramond (R) fields,
depending on how they behave under rotations around the origin in 
the punctured plane. 
NS fields are periodic and R fields are antiperiodic.  This means that the
operators in the NS sector form a closed algebra under operator product
expansions while the ones in the R sector do not. So if we project out the
R fields we obtain a consistent local quantum field theory, which will be
manifestly supersymmetric. Another way of
obtaining a consistent local theory is by projecting out the fermions,
which is usually called the GSO projection. This way we obtain what is 
usually called the spin model associated to the SCFT.
The important observation is that for the TIM 
we can associate each of the LG actions (\ref{slg}), (\ref{lg})
to each of these local projections. 

As it was argued in \cite{zam89}, a minimal model
perturbed by the $\hat\phi_{(1,3)}$ operator gives an
integrable theory. In the bosonic description (\ref{lg}) the 
operator $\hat\phi_{(1,3)}$
corresponds to $:\phi^6:$, and in the manifestly supersymmetric description
(\ref{slg}) to the auxiliary field $F=\int d^2 \theta \, \Phi$.
In both descriptions the LG action can still be used off criticality
as a guide to the solitonic structure of these massive deformations. 
An $S$-matrix based on this deformation of (\ref{lg}) 
has been proposed in \cite{zamtim}.
In this paper we study the $S$-matrix proposed in \cite{fendley}, corresponding
to the perturbed action
\be
S=\int d^2z \, d^2 \theta \, \left[{1 \over 2} (D\Phi)^2 + \Phi^3 +
\lambda \Phi\right]\ , \label{slgF}
\ee
which is manifestly supersymmetric\footnote{From now on we will refer to this 
perturbed model as TIM.}.
By looking at (\ref{slgF}) in terms of components it can be argued that
in this model there are solitonic and antisolitonic 
supersymmetric doublets, which we will denote by $(B,F)$ and
$(\bar{B},\bar{F})$.

Due to the fact that multi-soliton configurations have to alternate 
solitons and anti-solitons, there is an adjacency condition to be 
respected by the allowed multi-particle states.

In this paper we study the factorized scattering theory associated to the 
action (\ref{slgF}) in the presence of a boundary.
Some related work has been done by Chim in \cite{chim};
he solved the boundary Yang-Baxter equation (BYBE) 
for the $S$-matrix proposed in \cite{zamtim}, where 
supersymmetry acts non-locally. 
In that case it is difficult to identify which reflection
matrix corresponds to boundary supersymmetry preserving interactions. As we
will see, in the present formulation this identification is done in a 
transparent way, due to the fact that we are dealing with a manifestly
supersymmetric theory, where the supersymmetric charges act locally.

This paper is organized as follows. In the next section
we discuss general aspects of two-dimensional integrable theories. 
In section 3 we discuss the $S$-matrix proposed by Fendley to describe 
the $\hat\phi_{(1,3)}$ perturbation of the TIM. 
In section 4 we find the reflection matrices that preserve boundary 
supersymmetry. 
In section 5 we solve the boundary Yang-Baxter equation, and find 
the general reflection matrices which 
preserve integrability but not necessarily supersymmetry,
and relate these solutions to the solutions of section 4 in the appropriate
limits. 
In section 6 we discuss the possible boundary interactions which preserve 
integrability and supersymmetry, and connect them with the reflections 
matrices we have obtained.
In the last section we discuss our results and some possible extensions of 
this work.

\resection{Generalities about the Scattering Matrix}

In this section we briefly review the main aspects of factorized 
scattering theories that will be needed in the rest of the paper. 
This section is meant mainly to set the notation. As a general reference
for bulk integrable field theories in two-dimensions we refer the reader
to \cite{zamzam}.

We parameterize asymptotic states in terms of the rapidity 
variable $\t$, such that energy and momentum are given
by $p_0= m\cosh\t$ and $p_1=m\sinh\t$, respectively. One-particle states are
labeled by $|A_i(\t)\rangle_{in,out}$, where $A_i$ could be a boson
or a fermion. Since we have
solitons and anti-solitons, both bosonic and fermionic, we will denote 
solitons by $A_i$ and antisolitons by $\bar{A}_i$.
Multiparticle states are given by
$|A_{i_1}(\t_1)  A_{i_2}(\t_2) \ldots A_{i_n}(\t_n)\rangle_{in,out}$ 
such that $\t_1 > \t_2  > \ldots > \t_n$
for in-states, and the other way around for out-states. 
As a basis for one-particle
states we use $\{|B\rangle,|F\rangle, |\bar{B}\rangle,
|\bar{F}\rangle\}$ and for two-particle states we use
$\{|B\bar{B}\rangle, |F\bar{F}\rangle, |B\bar{F}\rangle, |F\bar{B}\rangle \}$.
The $S$-matrix is defined by
\be
|A_{i_1}(\t_1)  A_{i_2}(\t_2)\rangle_{in}=
S_{i_1 i_2}^{j_1 j_2}(\t_1-\t_2)|A_{j_2}(\t_2)  A_{j_1}(\t_1)\rangle_{out} \ ,
\ee
and is represented  graphically \footnote{Inside figures we will
denote the one-particle state $A_i$ simply by $i$.} in figure 1.
\vskip 0.5cm
\centerline{\epsffile{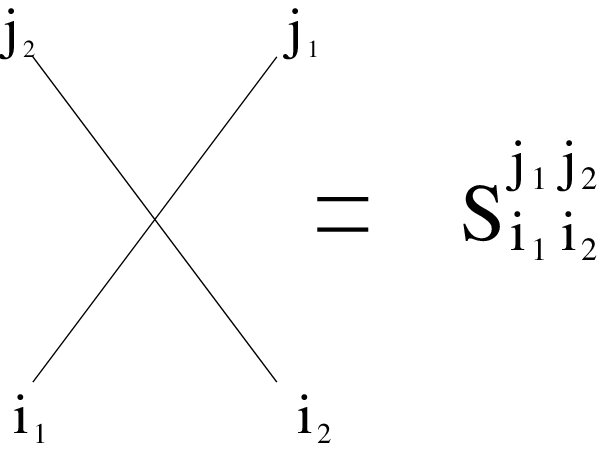}}
\vskip 0.5cm
\centerline{Fig. 1 The $S$-matrix.} 

Once we have a bulk $S$-matrix we can consider the associated problem
of ``boundarizing'' this model \cite{gzam,fk}.
A boundary scattering theory is described in the bulk by the same 
$S$-matrix as the bulk model we are studying. In order to have a complete
description we have to introduce the boundary
scattering matrix which tells us how particles scatter off the boundary.
This is the reflection matrix and is defined by
\be
|A_i(\t)\rangle = R_i^j(\t)|A_j(-\t)\rangle
\ee
and in a similar fashion as the bulk $S$-matrix, is represented graphically
as shown in figure 2.
\vskip 0.5cm
\centerline{\epsffile{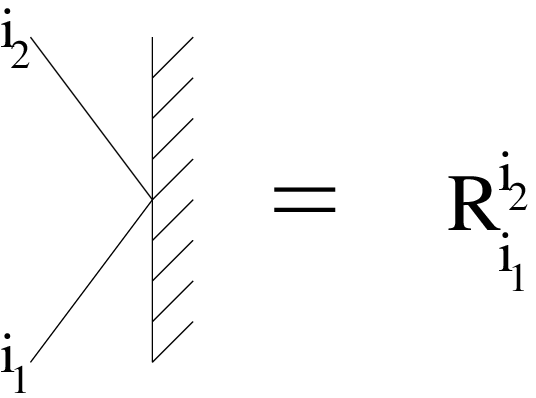}}
\vskip 0.5cm
\centerline{Fig. 2 The Reflection matrix.}

The consistency between boundary integrability and bulk integrability
is encoded in the boundary Yang-Baxter equation (BYBE) \cite{cher} which reads
\be
S_{i_1i_2}^{c_1c_2}(\t_{12}) R_{c_1}^{d_1}(\t_1)
S_{c_2d_1}^{d_2j_1}(\bar{\t}_{12}) R_{d_2}^{j_2}(\t_2)=
R_{i_2}^{c_2}(\t_2) S_{i_1c_2}^{c_1d_2}(\bar{\t}_{12}) R_{c_1}^{d_1}(\t_1)
S_{d_2d_1}^{j_2j_1}(\t_{12}) \ , \
\ee
where $\t_{12}=\t_1-\t_2$ and $\bar{\t}_{12}=\t_1+\t_2$.
The graphic representation of the BYBE is given in figure 3.
\vskip 0.5cm
\centerline{\epsffile{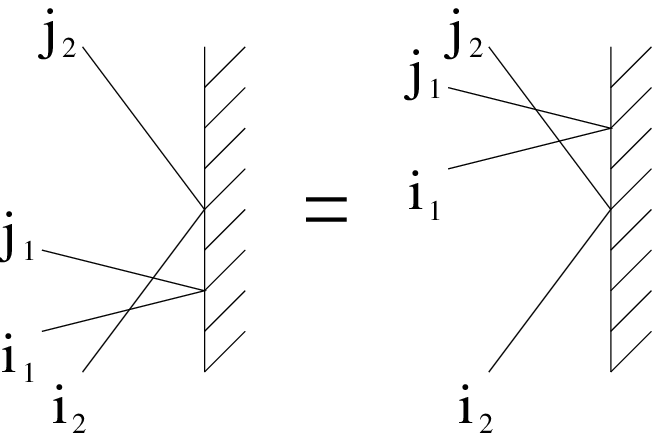}}
\vskip 0.5cm
\centerline{Fig. 3 The BYBE}
\vskip .3cm

\subsection{Supersymmetry Algebra}

The superfield sector of the TIM carries a representation of a
$N=1$ supersymmetry algebra with topological charge, which is given by
two supersymmetry generators, $Q_+$ and $Q_-$, and a fermion
number operator $Q_L$, whose eigenvalues measure if 
the state is bosonic ($+1$) or
fermionic ($-1$). The algebra reads explicitly \cite{wo}
\bea 
&&Q_+{}^2=p_0+p_1 \ , \qquad Q_-{}^2=p_0-p_1 \ , \nonumber \\
&& \qquad \qquad \{Q_+,Q_-\}=2 T \ , \label{susyalg} \\
&& \qquad \qquad \{Q_L,Q_{\pm}\}=0 \ , \nonumber 
\eea
and $T$ is the topological charge.
In our one-particle basis this algebra has the following realization:
\be
Q_+(\theta)= \sqrt{m} \, e^{\t \over 2}
\left(\begin{array}{cccc}
      0 & 1 & 0 & 0 \\
      1 & 0 & 0 & 0 \\
      0 & 0 & 0 & 1 \\
      0 & 0 & 1 & 0
      \end{array}\right), \ \ 
Q_-(\theta)= \sqrt{m} \, e^{-{\t \over 2}}
\left(\begin{array}{cccc}
      0 & {\rm e}^{i \alpha} & 0 & 0   \\
      {\rm e}^{-i \alpha} & 0 & 0 & 0  \\
      0 & 0 & 0 & -{\rm e}^{-i \alpha} \\
      0 & 0 & -{\rm e}^{i \alpha} & 0
      \end{array}\right).
\ee
The topological charge $T$ in (\ref{susyalg}) is given by 
\be
T= m \cos\alpha 
\left(\begin{array}{cccc}
      1 & 0 & 0 & 0 \\
      0 & 1 & 0 & 0 \\
      0 & 0 & -1 & 0 \\
      0 & 0 & 0 & -1
      \end{array}\right)\ . 
\ee
Notice that in our notation the Bogomolnyi bound 
$|T|\leq m $ is saturated when $\a=0$,
whereas the topological charge vanishes when $\alpha=\pi/2$.
Finally, the action of the supercharges on multi-particle 
states is
\be
\widehat{Q}_+(\t)=\sum_{l=1}^{N}Q_{+}^l(\t)\ ,
\quad \widehat{Q}_-(\t)=\sum_{l=1}^{N} Q_{-}^l(\t)\ ,
\ee
where $Q_{\pm }^l(\t)$ is defined by
\bea
&&Q_{\pm}^l(\t)|A_{a_1}(\t_1)\dots A_{a_N}(\t_N)\rangle= \nonumber \\
&&=|(Q_LA)_{a_1}(\t_1)\dots (Q_LA)_{a_{l-1}}(\t_{l-1})
(Q_{\pm}A)_{a_l}(\t_l) A_{a_{l+1}}(\t_{l+1}) \dots A_{a_{N}}(\t_{N}) 
\rangle \ .
\eea
This is a local realization of the $N=1$ supersymmetry algebra,
that is, the supersymmetric charges act on particle states,
in contrast to the non-local realization used in \cite{zamtim}.

\resection{Fendley's $S$-matrix}

In this section we briefly review the $S$-matrix proposed in \cite{fendley} 
to describe the TIM, and we refer the reader to that paper for a more 
detailed discussion.

In the two-particle basis given in section 2,
the $S$-matrix has the following form
\be
S(\t)=\left(
\begin{array}{cccc}
a(\t) & -b(\t) & 0 & 0 \\
b(\t) & a(\t) & 0 & 0 \\
0 & 0 & d(\t) & c(\t) \\
0 & 0 & -c(\t) & d(\t)
\end{array}
\right)\ .
\ee
The scattering theory turns out to be different if the 
Bogomolnyi bound is saturated ($\alpha=0$) or not ($\alpha \neq 0$). 
In this paper we will refer to these cases simply as saturated and 
non-saturated.
The $S$-matrix does not depend explicitly on $\alpha$, the
difference between the two cases being that in the former the amplitude
$c(\t)$ does not vanish, whereas in the latter $c(\t)=0$. This is due to
the fact that the $S$-matrix is determined by local interactions,
while the topological charge is a global property of the theory.

Commutativity with the supercharges implies that the amplitudes in the 
$S$-matrix are related by
\bea
b(\t)&=&a(\t)\sinh\frac{\t}{2} + c(\t) \cosh\frac{\t}{2} \ , \\
d(\t)&=&a(\t)\cosh\frac{\t}{2} + c(\t) \sinh\frac{\t}{2} \ . 
\label{bd}
\eea
{}From this expression we see immediately that this $S$-matrix satisfies the
free-fermion condition $a^2+b^2=c^2+d^2$,
which is extremely important for thermodynamical calculations 
\cite{fendley,MS1,fi,ahnsff}.
Crossing-symmetry requires $a(i\pi -\t)=a(\t)$ 
and $c(i\pi-\t)=-c(\t)$\footnote{Notice that the minus sign in the 
crossing relation
for $c(\t)$ is related to the fact that this is an amplitude involving
one fermion in the in-state.}, and from (\ref{bd}) one 
finds $b(i\pi-\t)=i d(\t)$ and 
$d(i\pi -\t)= - i b(\t)$. 

The Yang-Baxter equation (YBE) yields
\be
\frac{c}{a}=\e \tanh \t\ , 
\label{bulkyb}
\ee
where $\e=\pm 1$, and one finds
\be
\frac{b}{a}=\frac{\sinh p\t}{\cosh \t} \ ,\quad \frac{d}{a}(\t)=
\frac{\cosh p\t}{\cosh \t} \ ,
\ee 
where $p=3/2$ when $\e=1$ and  $p=-1/2$ when $\e=-1$.
The YBE implies also that scattering amplitudes should be the same
when exchanging solitons $\leftrightarrow$ antisolitons.

\noindent Finally, by solving unitarity it is found for the saturated case
\bea
a(\t)&=&2 \sinh ^2 (\frac{i\pi}{4}+\frac{\t}{2})
\prod_{l=0}^{\infty}
\frac{ \Gamma(\frac{3}{2}+l) \Gamma(\frac{1}{2}+l-\frac{i\t}{2\pi})
\Gamma(1+l+\frac{i\t}{2\pi})}
{ \Gamma(\frac{1}{2}+l) \Gamma(\frac{3}{2}+l+\frac{i\t}{2\pi})
\Gamma(1+l-\frac{i\t}{2\pi})} \times \nonumber \\
&&\times \frac{ \Gamma(\frac{1}{2}+2ql+2q)
\Gamma(\frac{1}{2}+ 2ql -\frac{iq\t}{\pi}) 
\Gamma(\frac{1}{2}+2ql+q+\frac{iq\t}{\pi})}
{\Gamma(\frac{1}{2}+2ql)\Gamma(\frac{1}{2}+2ql+2q + \frac{iq\t}{\pi})
\Gamma(\frac{1}{2}+2ql+q-\frac{iq\t}{\pi})}\ ,
\label{atheta}
\eea
where $q=1+|p|$.
Notice that the formula in \cite{fendley} differs from (\ref{atheta}) 
by a factor of $\tanh^2(\frac{i\pi}{4}+\frac{\t}{2})$. The reason for this
change is that $a(\t)$ in \cite{fendley} has a zero in the physical strip,
and therefore should not be taken as the minimal solution.
The integral representation for (\ref{atheta}) is
\be
a(\t)=\exp\left[-\int^{\infty}_{0}\frac{dt}{t}\, h(t)
                 \frac{\sin t\t \sin t(i\pi-\t)}{\cosh \pi t}\right]\ , 
\label{aint}
\ee
where 
\be
h(t)=
\frac{2}{\sinh \pi t}-
\frac{1}{\sinh 2 \pi t}-
\frac{1}{\sinh\frac{\pi t}{q}}\ . \label{ht}
\ee
As we have seen, the non-saturated case is gotten by setting $c(\t)=0$. 
This changes the unitarity equation and as a consequence the
minimal solution is
\be
a(\t)=\prod_{j=0}^{\infty}\left[\frac{
\Gamma(\frac{3}{2}+j)\Gamma(1+j+\frac{i\t}{2\pi})\Gamma(\frac{1}{2}+j-
\frac{i\t}{2\pi})}
{\Gamma(\frac{1}{2}+j)\Gamma(1+j-\frac{i\t}{2\pi})\Gamma(\frac{3}{2}+j+
\frac{i\t}{2\pi})}
\right]^2\ .
\ee
The integral representation is given by (\ref{aint}) with
\be
h(t)=\frac{2}{\sinh 2 \pi t}\ .
\ee

In the next sections we discuss the boundary scattering associated to this
$S$-matrix. Initially we will consider the reflection matrices for 
supersymmetry-preserving boundary interactions and later more general
solutions.

\resection{Supersymmetry Preserving Reflection Matrices}

The introduction of a boundary will
in general destroy some of the conserved charges. In particular
we can not preserve the whole supersymmetry of the bulk model, 
the best we can do being to preserve ``half'' of supersymmetry \cite{warner}. 
In this section we study which combination of the supercharges
can be preserved in the presence of a boundary and
the corresponding reflection matrices.

In the one-particle basis of section 2 we see that
all possible reflection processes can be encoded in a $4 \times 4$ matrix.
This reflection matrix can be written in general as
\be
{\cal R}(\t)=\left(
\begin{array}{cccc}
R & U \\
V & \bar R
\end{array}
\right)\ ,
\ee
where $R$, $\bar R$, $U$ and $V$ are $2 \times 2$  matrices. Our
convention is that rows label in-states and columns out-states.
Matrices $R$ and
$\bar R$ describe topological charge preserving processes, while  $U$ and $V$
describe reflections of solitons into antisolitons and vice-versa. 
Since the bulk $S$-matrix
satisfies an adjacency condition we should set $U=V=0$.
The reflection matrix has, therefore, block-diagonal form. 

$R$ and $\bar R$ can be written quite generally as
\be
R(\t)=\left(
\begin{array}{cccc}
R_b & P \\
Q &  R_f 
\end{array}
\right)\ , \quad 
\bar{R}(\t)=\left(
\begin{array}{cccc}
\bar R_b & \bar P \\
\bar Q &  \bar R_f 
\end{array}
\right)\ . \label{Rmatrix}
\ee
Non-diagonal amplitudes in (\ref{Rmatrix}) correspond to 
fermion-number changing processes.
We will see later that 
$R$ and $\bar R$ are connected by a simple transformation. 

\subsection{Boundary Supersymmetry}

We start by assuming that the boundary action\footnote{We speak of
a symbolic action.}
preserves both integrability and supersymmetry. As explained in 
\cite{warner,MS2}
only a linear combination of the supersymmetric charges can survive in
the presence of a boundary. It is easy to see that the only candidates
are $\widetilde{\cal Q}_{\pm}=Q_+ \pm Q_-$, since when squared these are  
the only linear combinations
which do not depend on linear momentum. We then require that the 
reflection matrix ``commutes'' with this new charge, that is
\be
\widetilde{\cal Q}_{\pm}(\t) {\cal{R}}(\t)=
{\cal{R}}(\t)\widetilde{\cal Q}_{\pm}(-\t) \ . \
\label{comm}
\ee
{}From this equation it is easy to see that $R(\t)$ should be of the form
\be
R_{\pm}(\t)=R_0(\t) \left(\begin{array}{cc}
\cosh (\frac{i\a}{2}^{\pm}-\frac{\t}{2}) & 
e^{ {{{i\a}\over 2}^{\pm}}}p(\t) \\
e^{-{{{i\a}\over 2}^{\pm}}}p(\t) & 
\cosh (\frac{i\a}{2}^{\pm}+\frac{\t}{2})
\end{array}
\right)\ ,
\ee
where $\a^+=\a$ and $\a^-=\a+\pi$,
and $\bar R_{\pm} (\t)$ of the form
\be
\bar R_{\pm}(\t)=\bar R_0(\t) \left(\begin{array}{cc}
\cosh (\frac{i\a}{2}^{\mp}+\frac{\t}{2}) & 
e^{- {{{i\a}\over 2}^{\mp}}}p(\t) \\
e^{{{{i\a}\over 2}^{\mp}}}p(\t) & 
\cosh (\frac{i\a}{2}^{\mp}-\frac{\t}{2})
\end{array}
\right)\ .
\ee
For convenience we have denoted by $R_{\pm}(\t)$ the reflection matrix 
which commute with the combinations $\widetilde{\cal Q}_{\pm}(\t)$ 
and we will adhere to this convention in the following.
It is evident that $\bar R$ can be obtained from $R$ by the simple substitution
\be
\a^{\pm} \rightarrow -\a^{\mp}\ , 
\label{bar}
\ee
as it can also be seen directly from the structure of the supercharges.
{}From now on we will concentrate only on $R$.

As a result, we see that the requirement of commutativity (\ref{comm}) 
determines the ratios between the diagonal elements, $R_f/R_b$, and between 
the off-diagonal ones, $Y=Q/P$, and in addition, it implies a 
precise relation 
between the reflection amplitudes in the solitonic and in the anti-solitonic 
sector.
In order to fix the last unknown ratio $p \equiv e^{-i\a^{\pm}/2}P/R_0$, 
we use the BYBE.
The relevant equation is the one corresponding to 
$B\bar B\rightarrow F \bar B$, 
and it reads explicitly
\bea
&&P(\t_1)\frac{R_b(\t_2)}{R_b(\t_1)} \left( x(\bar\t_{12}) - 
x(\t_{12}) \right) + Q(\t_1) \frac{R_b(\t_2)}{R_b(\t_1)} \left( y(\t_{12}) v(\bar\t_{12}) - v(\t_{12}) y(\bar\t_{12}))
\right)  = \nonumber \\
&&= P(\t_2) \left( y(\t_{12}) x(\bar\t_{12}) +
\frac{R_f(\t_1)}{R_b(\t_1)} x(\t_{12}) y(\bar\t_{12}) \right) +
Q(\t_2) \left( v(\bar\t_{12}) + \frac{R_f(\t_1)}{R_b(\t_1)} v(\t_{12}) 
\right)\ . \label{bybe1}
\eea
where $v=b/a$, $x=c/a$, and $y=d/a$.
To make the discussion clear let us treat
the saturated and non-saturated cases separately. 

\begin{itemize} 
\item{{\em{The Saturated Case}} ($c(\t)\neq0$)}

We find
\bea 
R_+(\t)&=&R_0(\t)\left(\begin{array}{cc}
                     1 & A \sinh q\t \\
           A \sinh q\t & 1
                     \end{array}\right), \label{rsat+}  \\[2mm]
R_-(\t)&=&R_0(\t)\left(\begin{array}{cc}
                     1 & A \cosh q\t \\
          -A \cosh q\t & -1
                     \end{array}\right), \label{rsat-}
\eea
where $A$ is a constant (which could be zero, of course).

\item{{\em{The Non-Saturated Case}} ($c(\t)=0$)}

We find 
\be 
R_{\pm}(\t)=R_0(\t)\left(
\begin{array}{cc}
\cosh(\frac{i\a}{2}^{\pm}-\frac{\t}{2}) & e^{ \frac{i\a}{2}^{\pm} } 
A\sinh \t \\
e^{-\frac{i\a}{2}^{\pm} } A\sinh \t & \cosh(\frac{i\a}{2}^{\pm}+\frac{\t}{2})
\end{array}
\right). \label{rnsat}
\ee
\end{itemize}
The reflection matrices depend explicitly on the topological charge,
as expected, since the introduction of a boundary brings global properties
into the local description of scattering theory.
All we will have to do now is to fix the overall 
prefactor by requiring unitarity and boundary crossing-unitarity. 
We will treat these requirements in the next subsection. 

\subsection{Unitarity and Boundary Crossing-Symmetry}

In this section we fix the prefactor $R_0(\t)$ in the following way.
As customary \cite{gzam}, 
we write $R_0(\t)=Z_1(\t)Z_2(\t)$ where $Z_1(\t)$ solves 
unitarity and
does nothing to boundary crossing-unitarity and  $Z_2(\t)$ solves
boundary crossing-unitarity and does nothing to unitarity. We restrict
ourselves to the minimal solutions, with no poles in the physical strip.
In the following we will also give integral representations for these
prefactors, since they are very useful for thermodynamical computations.

The unitarity requirement for the reflection matrix is given by
$R(\t)R(-\t)=1$, which in our case implies the following four equations:
\bea
R_b(\t)R_b(-\t)+P(\t)Q(-\t)&=&1\ , \nonumber \\
R_b(\t)P(-\t)+P(\t)R_f(-\t)&=&0\ , \nonumber \\
R_f(\t)R_f(-\t)+Q(\t)P(-\t)&=&1\ , \nonumber \\
R_f(\t)Q(-\t)+Q(\t)R_b(-\t)&=&0\ . \nonumber
\eea
It turns out that the the second and the fourth equation are automatically 
satisfied by (\ref{rsat+}), (\ref{rsat-}) and (\ref{rnsat}), whereas the 
first and the third equations are non-trivial. Let us discuss the 
saturated and non-saturated cases separately.

\begin{itemize}
\item{{\em{The Saturated Case}} ($c(\t)\neq 0$)}

The unitarity equation can be written as follows:
\be
Z_1(\t)Z_1(-\t)=\frac{A^{-2}}{\sinh (\kappa - q\t) \sinh (\kappa + q\t)}\ ,
\label{szeta1}
\ee
where the parameter $\kappa$ is defined by $\sinh \kappa =1/A$ 
for the ``$+$'' combination of supercharges 
and $\cosh \kappa =1/A$ for the ``$-$'' combination.
The minimal solutions are
$Z_1(\t) = \sigma(x,\t)$ and $Z_1(\t) = \sigma(x,\t)/\tanh^2 \kappa$,
respectively, $x=\frac{\pi}{2} + i\kappa$. The explicit expression 
of $\sigma$ as an infinite product of gamma functions \cite{gzam} is
\bea
\sigma(x,\t)&=&\frac{
\Pi(x,  \frac{\pi}{2}+i\t)\Pi(-x, \frac{\pi}{2}+i\t)\Pi(x, -\frac{\pi}{2}-i\t)
\Pi(-x,-\frac{\pi}{2}-i\t)}
{\Pi(x, \frac{\pi}{2})\Pi(-x, \frac{\pi}{2})\Pi(x, -\frac{\pi}{2})
\Pi(-x,-\frac{\pi}{2})}\ , \nonumber\\
\Pi(x,\t)&=&\prod_{l=0}^{\infty}
\frac{\Gamma(\frac{1}{2}+(2l+ \frac{1}{2})q+\frac{x}{\pi}+iq\frac{\t}{\pi})
      \Gamma(\frac{1}{2}+(2l+\frac{3}{2})q+\frac{x}{\pi})}
     {\Gamma(\frac{1}{2}+(2l+\frac{3}{2})q+\frac{x}{\pi}+iq\frac{\t}{\pi})
      \Gamma(\frac{1}{2}+(2l+\frac{1}{2})q+\frac{x}{\pi})}\ . 
\label{sigma}
\eea
The integral representation is given by
\be
Z_1(\t)=\exp\left[-{1\over 2}
\int^{\infty}_{0}\frac{dt}{t}\, 
\frac{ \cosh(\frac{\pi}{2}+ i\kappa) \frac{2t}{q} } 
{ \cosh \pi t \sinh \frac{\pi t}{q} }
\sin t \t
\sin t (i\pi-\t) \right] \ .
\ee
The last thing to be done is to impose boundary crossing-unitarity and
we will have the complete (minimal) reflection matrix. The boundary 
crossing-unitarity was introduced in \cite{gzam},
and we should note that their formula (3.35) assumes that one is dealing
with a parity preserving, neutral theory.
Since in our case we have fermions, we have to
pay attention to possible minus signs and charge conjugation phases (we refer
the reader to the appendix for a discussion on these issues).
The crossing-unitarity equation turns out to be
\be
K^{ab}(\t)=S_{a'b'}^{ba}(2\t)K^{a'b'}(-\t)\ ,
\label{bcross-u}
\ee
and in our case it reads
\be
Z_2(\frac{i\pi}{2}-\t)
=Z_2(\frac{i\pi}{2}+\t) \frac{2 a(2\t)}{\cosh 2\t} \cosh(
\frac{\eta\pi i}{4} + \frac{\t}{2})\cosh(\frac{\eta\pi i}{4}\e + q\t)\ ,
\label{satcross-u}
\ee
and $\eta$ is the sign of the charge combination in (\ref{comm}).
The minimal solution can be found by elementary methods and is given 
by
\be
Z_2(\t)=\frac{\cos(\frac{\pi}{8}-\frac{i\t}{2})}
{\cos(\frac{\pi}{8}+\frac{i\t}{2})} \Omega_1(\t)\Omega_2(\t) \ ,
\ee
where
\bea
&&\Omega_1(\t)=\prod_{k=0}^{\infty}
\frac{\Gamma(\frac{3}{4}+k-\frac{\eta}{4}+\frac{i\t}{2\pi})
             \Gamma(\frac{5}{4}+k+\frac{\eta}{4}+\frac{i\t}{2\pi})}     
            {\Gamma(\frac{3}{4}+k-\frac{\eta}{4}-\frac{i\t}{2\pi})
             \Gamma(\frac{5}{4}+k+\frac{\eta}{4}-\frac{i\t}{2\pi})} \times \nonumber \\
&&\quad\quad\quad\quad\quad\   \frac{\Gamma(1+k+\frac{\eta}{4}-\frac{i\t}{2\pi})
             \Gamma(1+k-\frac{\eta}{4}-\frac{i\t}{2\pi})}
            {\Gamma(1+k+\frac{\eta}{4}+\frac{i\t}{2\pi})
             \Gamma(1+k-\frac{\eta}{4}+\frac{i\t}{2\pi})} \ ,\\
&&\Omega_1(\t)=\prod_{k=0}^{\infty}
\frac{\Gamma(\frac{1}{2}+2q(k+\frac{3}{4})+\frac{\eta\e}{4}+\frac{iq\t}{\pi})
      \Gamma(\frac{1}{2}+2q(k+\frac{1}{4})-\frac{\eta\e}{4}+\frac{iq\t}{\pi})}
      {\Gamma(\frac{1}{2}+2q(k+\frac{3}{4})+\frac{\eta\e}{4}-\frac{iq\t}{\pi})
      \Gamma(\frac{1}{2}+2q(k+\frac{1}{4})-\frac{\eta\e}{4}-\frac{iq\t}{\pi})} \times \nonumber \\
&&\quad\quad\quad\quad\quad\  
\frac{\Gamma(\frac{1}{2}+2q(k+\frac{1}{2})+\frac{\eta\e}{4}-\frac{iq\t}{\pi})
      \Gamma(\frac{1}{2}+2q(k+\frac{1}{2})-\frac{\eta\e}{4}-\frac{iq\t}{\pi})}
     {\Gamma(\frac{1}{2}+2q(k+\frac{1}{2})+\frac{\eta\e}{4}+\frac{iq\t}{\pi})
      \Gamma(\frac{1}{2}+2q(k+\frac{1}{2})-\frac{\eta\e}{4}+\frac{iq\t}{\pi})}
\ .
\eea
The integral expression for $Z_2(\t)$ is
\be
Z_2(\t) 
=\exp\left[-\frac{i}{2}\int_{0}^{\infty}\frac{dt}{t} 
\frac{\sin \t t}{\cosh\frac{\pi t}{4}
\cosh\frac{\pi t}{2}} \left( -1 + A(t) + B(t) \right) \right]\ , \label{szeta2}
\ee
where the functions $A(t)$ and $B(t)$ are given by
\be
A(t)=\frac{\sinh \frac{\pi t}{4}(1 + 2\eta)}
{\sinh \pi \t}\ , \quad
B(t)=\frac{\sinh \frac{\pi t}{4q}(q +\eta\e)}
{\sinh \frac{\pi t}{2q}}\ .
\ee
The prefactor $R_0(\t)$ can be easily written now. 
This concludes the discussion for the saturated case.

\item{{\em{The Non-Saturated Case}} ($c(\t)=0$)}

The unitarity condition is
\be
Z_1(\t)Z_1(-\t)=\left[
\cosh(\frac{i\a^{\pm}}{2}-\frac{\t}{2})
 \cosh(\frac{i\a^{\pm}}{2}+\frac{\t}{2})
-A^2\sinh^2\t \right]^{-1}\ .
\ee
The solution can be expressed in terms of the function 
$\sigma(x,\t)$ defined in (\ref{sigma}) with $q=1/2$ and it is given by
\be
Z_1(\t)=\frac{\sqrt{\cosh 2\phi-\cosh 2\xi}}{\sqrt 2 \sinh\phi \cosh\xi}
\s(\frac{\pi}{2} + i\phi,\t) \s(i\xi,\t)\ . \label{nszeta1}
\ee
where the parameter $\xi$, $\phi$ are defined by
\be
\cosh 2\phi -\cosh 2\xi= \frac{1}{2A^2}\  , \quad
\cosh 2\phi \cosh 2\xi= 1+\frac{\cosh i\a^{\pm}}{2A^2} \ .
\ee
Finally boundary crossing-symmetry yields the following equation
for $Z_2(\t)$:
\be
Z_2(\frac{i\pi}{2}-\t)
=Z_2(\frac{i\pi}{2}+\t) a(2\t) \cosh \t\ ,
\label{nsatcross-u}
\ee
whose minimal solution is given by
\be
Z_2(\t)=\prod_{l=0}^{\infty}\frac{\Gamma(1+l+\frac{i\t}{\pi})
\Gamma^2(\frac{3}{2}+2l-\frac{i\t}{\pi})}
{\Gamma(1+l-\frac{i\t}{\pi})\Gamma^2(\frac{3}{2}+2l+\frac{i\t}{\pi})}\ .
\ee
This has a simple integral expression which we quote below:
\be
Z_2(\t)=\exp\left[-i \int_0^\infty \frac{dt}{t}
{{\sin t\t \sinh\frac{\pi t}{4}} \over
\sinh \pi t \cosh \frac{\pi t}{4}} \right] \ . \label{nszeta2}
\ee
\end{itemize}
The complete minimal solution for the prefactor $R_0(\t)$ 
can be easily written now.
This concludes our description of the supersymmetry preserving boundary
reflection matrices.

\resection{General Reflection Matrices}

In the previous section we have computed the reflection amplitudes
assuming that the underlying boundary interaction preserves both
integrability and supersymmetry. This last requirement
simplified computations since the constraint (\ref{comm})
severely restricts the form of $R(\t)$.
In order to study the interplay between integrability and supersymmetry 
in the presence of a boundary \footnote{See \cite{inodzh,prata} 
for related discussions in the Lagrangian approach.}
it is also interesting to invert the logic of the previous section:
first require only boundary integrability, namely solve the full BYBE, and
then try to understand in what limits, if any, supersymmetry can be restored.
This is what we do this section.
As an initial remark, notice that since the bulk $S$-matrix
does not change under the substitution soliton $\leftrightarrow$ anti-soliton,
the reflection amplitudes in the soliton sector and in the anti-soliton
sector will satisfy the same BYBEs. As a consequence, the functional form
of the amplitude ratios are the same in the two sectors, but 
depending on two different sets of free parameters. 
This means that we can again concentrate only on
$R(\t)$. Clearly enough, we expect that the two sets of parameters have to 
be related somehow, in order to recover boundary supersymmetry,
as in (\ref{bar}).

Let us look initially at the BYBE corresponding to the factorization of
$B \bar{B} \rightarrow F \bar{F}$ reflection process
\be
\frac{R_f(\t_2)}{R_b(\t_2)}v(\bar \t_{12}) + 
\frac{R_f(\t_1)}{R_b(\t_1)}\frac{R_f(\t_2)}{R_b(\t_2)} v(\t_{12})=v(\t_{12}) +
\frac{R_f(\t_1)}{R_b(\t_1)} v(\bar\t_{12})\ ,
\label{dbyb}
\ee
This equation has the interesting (and simplifying) feature of 
being the same whether we impose that there are non-diagonal processes 
or not. The simplest solution of (\ref{dbyb}) is $R_f(\t)/R_b(\t)=\pm 1$, 
which is a solution for any $v(\t)$.
In order to obtain other solutions for (\ref{dbyb}) we convert it into a
differential equation for $f(\t)=R_f(\t)/R_b(\t)$:
\be
\frac{\dot{f}(\t)}{1-f^2(\t)}=-\frac{\dot{v}(0)}{v{(2\t)}} \ . \label{fdiff}
\ee
The solution of (\ref{fdiff}) is 
\be
f(\t)=\frac{R_f(\t)}{R_b(\t)}=- \tanh\left(F(\t) + k \right) \ ,
\ee
where $F(\t)=\int^\t d\t' \, \dot{v}(0)/v(2\t')$ and $k$ is an integration
constant. In the non-saturated case this gives 
\be
\frac{R_f(\t)}{R_b(\t)}=\frac{\cosh(\frac{i\a}{2}'+\frac{\t}{2})}
{\cosh(\frac{i\a}{2}'-\frac{\t}{2})} \ , 
\ee
where $\a'$ is such that
\be
\tanh \frac{i\a}{2}'=-e^{2k} \ .
\ee
In the saturated case the solutions of (\ref{fdiff}) have an ``unusual'' 
functional form and we will not treat them in the following. Nonetheless
it would be an interesting problem to analyze their physical meaning.

The ratio $r(\t)=Q(\t)/P(\t)$ is fixed by the BYBEs corresponding to 
the factorization of $B \bar B \rightarrow B\bar B$ and 
of $B \bar F \rightarrow B \bar F$:
\bea
(r(\t_2)-r(\t_1))y(\bar{\t}_{12})&=&(1-r(\t_1)r(\t_2))
x(\bar{\t}_{12})v(\t_{12}) \ , \label{bybe2} \\
(r(\t_2)-r(\t_1))y(\t_{12})&=&(1-r(\t_1)r(\t_2))
x(\t_{12})v(\bar{\t}_{12}) \ . \label{bybe3}
\eea
These equations imply that $r(\t)=\pm 1$ in the saturated case
and $r(\t)=\g= \mbox{constant}$, in the non-saturated case. 
Since the solution of unitarity and boundary crossing-unitarity
is closely related to the one in section 3.1 we will simply quote the results 
in the following.

\begin{itemize}
\item{{\em{The Saturated Case}} ($c(\t)\neq 0$)}

The ratio between the diagonal elements, as well as the ratio
between off-diagonal ones, is fixed to be $\pm 1$.
{}From (\ref{bybe1}) we find
\be
R(\t)=R_0(\t)\left(
\begin{array}{cc}
1 & A\sinh(1 \pm \e p)\t \\
\pm A\sinh(1 \pm \e p)\t & 1
\end{array}
\right)\ , 
\ee
and 
\be
R(\t)=R_0(\t)\left(
\begin{array}{cc}
1 & A\cosh(1 \mp \e p)\t \\
\pm A\cosh(1 \mp \e p)\t & -1
\end{array}   
\right)\ ,
\ee
where $\e$ is the sign of $c(\t)$ in (\ref{bulkyb}). The prefactor in 
this case is easily seen to be the same as in (\ref{szeta1}), 
(\ref{szeta2})
with the substitution $q \rightarrow \tilde{q}=1 \pm \epsilon p$
and $q \rightarrow \tilde{q}=1 \mp \epsilon p$, respectively.
Notice that out of four possible sign combinations, two coincide exactly
with (\ref{rsat+}, (\ref{rsat-}).

\item{{\em{The Non-Saturated Case}} ($c(\t)=0$)}

The ratio of the diagonal elements is exactly
the same as the boundary supersymmetry-preserving one, with an arbitrary
parameter $\a'$ which is not necessarily related to the topological charge.
The solution of (\ref{bybe2}), (\ref{bybe3}) fixes the ratio 
between the off-diagonal elements of the reflection matrix to be a constant. 
We get
\be
R(\t)=R_0(\t)\left(
\begin{array}{cc}
\cosh(\frac{i\a'}{2}-\frac{\t}{2}) & P(\t) \\
\g P(\t) & \cosh(\frac{i\a'}{2}+\frac{\t}{2})
\end{array}
\right) \ .
\ee
Only when $\g=\exp(-i\a')$ we recover
the supersymmetric solution. Therefore we can think of $\g$ as a parameter that
``measures'' how far we are from a supersymmetry preserving 
boundary interaction.
To fix $P(\t)$ we use (\ref{bybe1}) obtaining
$P(\t)=A \sinh\t$. Finally, the
prefactor $R_0(\t)$ is fixed to be the same as in (\ref{nszeta1}) and 
(\ref{nszeta2}) with parameters $\xi$ and $\phi$ defined now by
\be
\cosh 2\phi -\cosh 2\xi= \frac{\gamma}{2A^2}\  , \quad
\cosh 2\phi \cosh 2\xi= 1+\frac{\gamma\cosh i\a'}{2A^2} \ .
\ee
\end{itemize}
This concludes the analysis of the reflection matrices for the TIM. In
the next section we discuss the possible boundary perturbations connected
to these solutions.

\resection{Boundary Perturbations}

One of the main problems in boundary factorized scattering theory is that
it is very difficult, in general, to relate 
solutions of the BYBE to specific boundary perturbations
or, in other words, to connect the parameters appearing in the reflection 
matrices with the actual boundary coupling constants in a Lagrangian 
description, if the model admits one 
\footnote{for related discussions 
in the case of boundary sine-Gordon see \cite{gzam},
\cite{aklc}.}.
In this section we connect our reflection matrices 
to specific boundary perturbations, within the formalism of deformed 
boundary conformal field theory (BCFT) \cite{gzam}.
A microscopic analysis of conformal boundary conditions for the TIM
has been performed in \cite{chim}, where the correspondence with $A_4$
RSOS model has been used. In that formulation it is difficult to analyze
supersymmetric boundary conditions, whereas in the present case it is
quite natural.

Let us notice initially that in the massless bulk limit the topological charge 
in the supersymmetry algebra vanishes \cite{warner}, 
and as a consequence we do not 
expect any difference between the saturated and non-saturated case 
from the point of view of deformed CFT .

After this preliminary remark, recall that  
in the bulk our model can be formulated as the massive deformation of the 
NS sector of ${\cal SM}_3$ by the 
relevant primary operator $\hat\Phi_{(1,3)}$.
In fact, the chiral NS superfield has two components, the 
energy operator and the sub-leading (vacancy) operator, 
$\hat\Phi_{(1,3)}=\e+{\bf\t}\e'$ 
The perturbation by $\e'$ preserves supersymmetry and the one by $\e$ 
breaks it \cite{dm}. 

A boundary field theory is defined by specifying the 
conformal boundary conditions (CBCs) and the boundary perturbation. 
It is well known \cite{cardy324} that in minimal unitary (super)conformal 
models the CBCs are in one-to-one correspondence with primary 
operators. This means that in the NS sector of ${\cal SM}_3$ , the possible 
CBCs that do not break superconformal invariance correspond to 
the primary superfields $\hat\Phi_{(1,1)}$  and $\hat\Phi_{(1,3)}$. 
The CBCs determine the spectrum of allowed boundary operators 
\cite{cardy275,cardy324} 
and it turns out that in the first case the only boundary 
operators that can appear are the identity ${\bf 1}$ and 
the irrelevant operator $\e''$, whereas in the second $\e$ and $\e'$.

Once we know the boundary 
operator content of a BCFT we can study which ones will
preserve boundary integrability.
The argument of \cite{gzam} can be rephrased by saying \cite{warner} that a 
boundary operator preserves integrability if it is in 
the same representation 
of the relevant conformal algebra as the bulk perturbation. 
In our case this means that the perturbation by the boundary superfield 
$\hat\Phi_{(1,3)}$ is integrable; furthermore, the perturbation by $\e$ 
breaks supersymmetry while the perturbation by $\e'$ 
preserves it, as it can be easily verified to first order in 
conformal perturbation theory.
Let us notice, finally, that from Cardy's analysis \cite{cardy275}
(see also \cite{fro}) it turns out that the free boundary conditions do not 
support the boundary operators $\e$ and $\e'$. A reasonable proposal
is that the reflection matrices we obtained correspond to some sort of
fixed boundary conditions perturbed respectively by the operators 
$\e'$ and $\e$.

\resection{Discussion and Conclusions}

In this paper we have found the exact reflection matrices for the
$S$-matrix proposed by Fendley to describe the superfield sector of the
tricritical Ising model, where supersymmetry acts locally. 
Supersymmetry fixes, almost completely,
the structure of boundary scattering and predicts a universal ratio
for the amplitudes of bosons and fermions scattering diagonally 
off the boundary. More explicitly, the requirement of boundary 
supersymmetry alone fixes $R_b/R_f$,
$Q/P$ and establishes the precise relation between $R$ and $\bar R$.

We also solved the BYBE in general
and we showed that it fixes $R_b/R_f$,
$Q/P$ and that $R$ and $\bar R$ should have the same functional form.
We were able to connect some of theses
solutions to the supersymmetry preserving ones.

As a next step it would be interesting to compute correlation functions in this
realization of the TIM by means of the form-factor approach. A first
step in the computation of supersymmetric form-factors has been done in 
the paper \cite{ahnsff}.

As a last remark we should mention that a thermodynamical Bethe ansatz 
computation of 
finite-size effects would be very useful in order 
to confirm that this description in term of supersymmetric soliton doublet 
is the correct scattering theory for the massive excitations of the 
supersymmetric TIM.
In any case the necessity of introducing some CDD factors does not 
change the structure of our reflection matrices.

\section*{Acknowledgments}
We would like to thank G.~Mussardo for a
critical reading of the paper and for many valuable suggestions,
P.~Fendley for useful remarks concerning the $S$-matrix used in this 
paper and R.~Musto for useful comments.
One of us (MM) would like to thank K.~Schoutens for discussions on
related issues in an early stage of this work and Lis for
encouragement.

\newpage

\section*{Appendix}

In this appendix we discuss some subtleties in the boundary 
crossing-unitarity condition that arise in models with fermions.
These are well understood, but since we have not found
an explicit discussion in the literature about these 
issues, we will present it here.
Recall that the crossing symmetry property of the $S$-matrix can be 
written as 
\be
S_{ab}^{cd}(i\pi-\t)=C_{bb'}S^{b'c}_{d'a}(\t)C^{d'd} \ ,
\ee
where $C_{ab}$ is the charge conjugation matrix and $C^{ab}$ is its inverse.
{}From the crossing properties of our bulk $S$-matrix we find that the only 
non vanishing elements of the charge conjugation matrix
can be chosen in the following way: 
\be
\begin{array}{ll}
C_{B\bar B}=C_{\bar B B}=1,& C_{F \bar F}=C_{\bar F F}=i \ ,
\end{array}
\ee
As it is well known \cite{gzam}, 
the reflection amplitude $R(\t)$ is the
analytic continuation of the amplitude $K(\t)$ to the domain 
$\mbox{Im}\t=\frac{i\pi}{2}$, $\mbox{Re}\t<0$. 
In models that are not invariant under charge conjugation, 
the appropriate analytic continuation 
involves the charge conjugation matrix as follows:
\be
K^{ab}(\t)=C^{aa'}R^{b}_{a'}(\frac{i\pi}{2}-\t) \ .
\ee
In order to write down the boundary crossing-unitarity equation,
without assuming we are dealing with a theory 
invariant under the usual discrete 
symmetries, we come back to the argument of Ghoshal-Zamolodchikov.
The amplitude $K^{ab}(\t)$ at positive real $\t$ is the coefficient of the 
two-particle contribution in the expansion of the boundary state in terms of 
out states,
\be
|B\rangle=\left[1+\int_0^{\infty}d\t K^{ab}(\t)A_a(-\t)A_b(\t)+\dots\right]
|0\rangle ,
\ee
whereas at negative real $\t$ it is interpreted as the two-particle 
contribution in the in states basis, 
\be
|B\rangle=\left[1+\int_0^{\infty}d\t K^{ab}(-\t)A_a(\t)A_b(-\t)+\dots\right]
|0\rangle .
\ee
The boundary cross-unitarity condition is obtained as a consistency condition
of these two expressions, using the fact that in and out states are 
related through the $S$-matrix:
\be
K^{ab}(\t)=S_{a'b'}^{ba}(2\t)K^{a'b'}(-\t).
\ee
Notice the different ordering of indices with respect to equation (3.35) in 
\cite{gzam}, where invariance under charge conjugation, parity and 
time reversal were assumed.

\end{document}